\documentclass[twocolumn]{aastex62}
\usepackage[utf8]{inputenc}
\usepackage{color}
\usepackage{graphicx}
\usepackage{amsmath}
\usepackage{bm}
\usepackage{acronym}
\usepackage{ifthen}
\usepackage{blindtext}
\usepackage[normalem]{ulem}
\usepackage{hyperref}
\usepackage{etoolbox}
\usepackage{fancyhdr}
\usepackage{xspace}
\usepackage{textcomp}
\usepackage{multirow}
\usepackage[caption=false]{subfig}
\usepackage{lineno}

\usepackage[utf8]{inputenc}

\renewcommand{\today}{\number\day\space\ifcase\month\or
  January\or February\or March\or April\or May\or June\or
  July\or August\or September\or October\or November\or December\fi
  \space\number\year}

\newcommand{\makevisible}[1]{\textbf{\color{cyan}#1}}
\newcommand{\switch}[1]{%
  \ifthenelse{\equal{#1}{0}}{\renewcommand{\makevisible}[1]{}}{}}
\switch{1}  

\newtoggle{fullauthorlist}
\toggletrue{fullauthorlist}

\usepackage[normalem]{ulem}
\def\gw{gravitational wave\xspace}
\def\gwh{gravitational-wave\xspace}

\newacro{S/N}{signal-to-noise ratio}

\newacro{PN}{post-Newtonian}

\newacro{O3}{the third observing run}

\newacro{O2}{the second observing run}

\newacro{MSP}{millisecond pulsar}
\def\MSP{\ac{MSP}\xspace}
\newacro{CW}{\emph{Continuous Wave}}
\def\CW{\ac{CW}\xspace}
\usepackage{soul}

\newcommand\scinum[2]{\ensuremath{#1\!\times\!10^{#2}}}

\definecolor{spring}{rgb}{0.7,0.9,0.7}
\definecolor{brick}{rgb}{0.7,0.2,0.1}
\definecolor{redHL}{rgb}{1.0,0.5,0.5}

\begin{document}

\reportnum{LIGO-P2000029}

\title{Gravitational-wave constraints on the equatorial ellipticity of millisecond pulsars}


%


\iftoggle{fullauthorlist}{
\author{R.~Abbott}
\affiliation{LIGO, California Institute of Technology, Pasadena, CA 91125, USA}
\author{T.~D.~Abbott}
\affiliation{Louisiana State University, Baton Rouge, LA 70803, USA}
\author{S.~Abraham}
\affiliation{Inter-University Centre for Astronomy and Astrophysics, Pune 411007, India}
\author{F.~Acernese}
\affiliation{Dipartimento di Farmacia, Universit\`a di Salerno, I-84084 Fisciano, Salerno, Italy  }
\affiliation{INFN, Sezione di Napoli, Complesso Universitario di Monte S.Angelo, I-80126 Napoli, Italy  }
\author{K.~Ackley}
\affiliation{OzGrav, School of Physics \& Astronomy, Monash University, Clayton 3800, Victoria, Australia}
\author{A.~Adams}
\affiliation{Christopher Newport University, Newport News, VA 23606, USA}
\author{C.~Adams}
\affiliation{LIGO Livingston Observatory, Livingston, LA 70754, USA}
\author{R.~X.~Adhikari}
\affiliation{LIGO, California Institute of Technology, Pasadena, CA 91125, USA}
\author{V.~B.~Adya}
\affiliation{OzGrav, Australian National University, Canberra, Australian Capital Territory 0200, Australia}
\author{C.~Affeldt}
\affiliation{Max Planck Institute for Gravitational Physics (Albert Einstein Institute), D-30167 Hannover, Germany}
\affiliation{Leibniz Universit\"at Hannover, D-30167 Hannover, Germany}
% ...

\collaboration{The LIGO Scientific Collaboration and the Virgo Collaboration}

  \author{M.~J.~Keith}
\affiliation{Jodrell Bank Centre for Astrophysics, School of Physics and Astronomy, University of Manchester, Manchester, M13 9PL, UK}
\author{A.~G.~Lyne}
\affiliation{Jodrell Bank Centre for Astrophysics, School of Physics and Astronomy, University of Manchester, Manchester, M13 9PL, UK}
\author{J.~Palfreyman}
\affiliation{Department of Physical Sciences, University of Tasmania, Private Bag 37, Hobart, Tasmania 7001, Australia}
\author{B.~Shaw}
\affiliation{Jodrell Bank Centre for Astrophysics, School of Physics and Astronomy, University of Manchester, Manchester, M13 9PL, UK}
\author{B.~W.~Stappers}
\affiliation{Jodrell Bank Centre for Astrophysics, School of Physics and Astronomy, University of Manchester, Manchester, M13 9PL, UK}
\author{P.~Weltevrede}
\affiliation{Jodrell Bank Centre for Astrophysics, School of Physics and Astronomy, University of Manchester, Manchester, M13 9PL, UK}

}
{
  \author{The LIGO Scientific Collaboration}
  \affiliation{LSC}
  \author{The Virgo Collaboration}
  \affiliation{Virgo}
}

\date{\today}

\begin{abstract}
We present a search for continuous gravitational waves from five radio pulsars, comprising three recycled pulsars (PSR~J0437\textminus4715, PSR~J0711\textminus6830, and PSR~J0737\textminus3039A) and two young pulsars: the Crab pulsar (J0534+2200) and the Vela pulsar (J0835\textminus4510). We use data from the third observing run of Advanced LIGO and Virgo combined with data from their first and second observing runs.
For the first time we are able to match (for PSR~J0437\textminus4715) or surpass (for PSR~J0711\textminus6830) the indirect limits on gravitational-wave emission from recycled pulsars inferred from their observed spin-downs, and constrain their equatorial ellipticities to be less than $10^{-8}$. 
For each of the five pulsars, we perform targeted searches that assume a tight coupling between the gravitational-wave and electromagnetic signal phase evolution.
We also present constraints on PSR~J0711\textminus6830, the Crab pulsar and the Vela pulsar from a search that relaxes this assumption, allowing the gravitational-wave signal to vary from the electromagnetic expectation within a narrow band of frequencies and frequency derivatives.

\end{abstract}



\section{Introduction} \label{sec:intro}

The field of \gwh astronomy is now firmly established, with the detection of multiple compact binary coalescences by the LIGO and Virgo observatories.  
These discoveries have included multiple black hole-black hole coalescences \citep{2019PhRvX...9c1040A}, and binary neutron star coalescences \citep{2017PhRvL.119p1101A, 2020arXiv200101761T}.  
Resulting studies have included tests of strong-field General Relativity \citep{2019PhRvD.100j4036A}, measurement of the Hubble parameter \citep{2017Natur.551...85A,2019ApJ...871L..13F,2019arXiv190806060T}, confirmation of the association between binary neutron star coalescence and short gamma ray bursts \citep{2017ApJ...848L..12A}, and information on the pressure-density relation for ultra-high density matter \citep{2018PhRvL.121p1101A}.

Other types of gravitational-waves sources, however, remain to be detected, including \CW sources. \acp{CW} have a relatively simple structure, consisting of just one or two harmonic components, whose amplitudes and frequencies change slowly on the year-long timescales of observations.  The prime candidates for producing such \CW signals are spinning neutron stars that have non-axisymmetric distortions, caused either by a solid deformation, probably sourced through some combination of elastic and magnetic stresses, or by the excitation of fluid modes of oscillation also referred as {\it r-modes\citep{2015MNRAS.446.3631A}}.  The astrophysical pay-off in making a detection would be considerable, shedding light on the structure of the star.
Moreover a \CW detection would allow further tests of general relativity, such as constraining non-standard \gwh polarizations \citep{isi2017probing}.  A recent review of the astrophysics of \CW sources is given in \citet{2018ASSL..457..673G}.

\subsection{Continuous wave searches}

\CW searches can be divided into three main types.  \emph{Targeted searches} look for signals from known pulsars whose rotational phase is accurately determined from electromagnetic observations, considerably simplifying the search.  \emph{Directed searches} look for signals from small sky areas, such as supernova remnants, where a neutron star is believed to reside, but for which no timing solution exists, so that a wide range of rotational parameters needs to be searched over.  \emph{All-sky} searches look for signals over all sky directions and also over a wide range of rotational parameters.  Many searches of these three types have already been carried out, using LIGO and Virgo data.  For recent examples, see \citet{2019ApJ...879...10A, 2019ApJ...875..122A, 2019PhRvD.100b4004A}.  No detections have been made, and consequently upper limits have been set on the strengths of such signals.

In this paper we report new results of targeted searches for \CW signals from five pulsars, using the most recent LIGO and Virgo data sets. Specifically, we use data from the first and second observing runs (O1 and O2), together with data from the first half of the third observing run (O3a), allowing us to set improved upper limits compared to other recent searches, e.g., \citet{2019ApJ...879...10A}.

It is possible to carry out such searches for many more (several hundred) known pulsars \citep{2019ApJ...879...10A}.  
We report results here for pulsars of particular interest.  
Specifically, we target three older, recycled pulsars, two of which are millisecond pulsars and one of which is only mildly recycled, that are believed to have undergone periods of accretion, and two very young pulsars: Crab and Vela. We search for the older pulsars, and particularly the recycled pulsars, because the signal amplitude is proportional to the square of the frequency, and therefore only small distortions are necessary to make a detection possible (see equation \eqref{eq:h0}).  The young pulsars are interesting because their rapid spin-down means that only a small fraction of their spin-down energy need go into the \gwh channel for a detection to be possible. Here we obtain direct \gwh observational limits that are at or below the spin-down limits for two of the recycled pulsars.  This is the first time the spin-down limit has been equalled or surpassed for a recycled pulsar.  As such, this represents a significant milestone for \gwh astronomy.

The structure of this paper is as follows.  In Section~\ref{sec:model} we describe the signal models we used.  
In Section~\ref{sec:methods} we discuss the analysis methods used in the searches.  
In Section~\ref{sec:data} we describe both the \gwh data we used, and also the radio pulsar data that was used to produce the timing solutions on which the \gwh searches were based.  In Section~\ref{sec:results} we describe our results, which are then discussed in Section~\ref{sec:discussion}.  Finally, in Section~\ref{sec:conclusions}, we draw some conclusions.

\subsection{Signal models}\label{sec:model}

We will assume \gwh emission that is tied closely to the rotational phase of the star.  
In the simplest case of a triaxial star spinning steadily about a principal moment of inertia axis, the \gwh emission is at exactly twice the star's spin frequency.  

There are several mechanisms, however, that can produce slightly different signals.  Free precession of the star can produce a  small frequency offset between the \gwh and (twice) the spin frequency, and also produce a lower harmonic, at or close to the spin frequency \citep{1979PhRvD..20..351Z, 2002MNRAS.331..203J}.  In most cases, free precession would modulate the observed radio pulsar frequency, a phenomenon not commonly observed in the pulsar population.  However, as noted by \citet{2010MNRAS.402.2503J}, the presence of a superfluid component within the star with a spin axis misaligned from that of the main rotation can produce this dual-harmonic emission, while leaving no imprint on the radio emission.  Another possibility is that the dominant \gwh emission is produced by a solid core \citep{glendenning_96, 2005PhRvL..95u1101O} whose spin frequency is slightly greater than that of the crust, again leading to a small mismatch between the gravitational and (twice) the radio pulsar frequency; see \citet{2008ApJ...683L..45A}.

With these considerations in mind, we follow previous \CW analyses and carry out three different sorts of search within this paper.  The simplest search assumes a single \gwh component, at exactly twice the observed spin frequency, as deduced from radio pulsar observations.  We carry out `dual harmonic searches', allowing for emission at both one and two times the spin frequency.  And we also carry out searches allowing for a small mismatch between the electromagnetic and gravitational signal frequencies, so-called ``narrowband'' searches.

The basic form of the waveform used in dual harmonic searches is described in detail in \citet{2015MNRAS.453...53J}, and used to perform searches in \citet{2015MNRAS.453.4399P}, and \citet{2017ApJ...839...12A, 2019ApJ...879...10A}.  We refer the reader to these papers, and in particular Section~1.1 and Appendix~A of \citet{2017ApJ...839...12A}.  We reproduce the main results here for completeness.  

If we denote the signals at one and two times the spin frequency as $h_{21}(t)$ and $h_{22}(t)$, respectively, we have
\begin{eqnarray}
\nonumber
h_{21} &= - \frac{C_{21}}{2} \Big[&F^D_+(\alpha, \delta, \psi; t) \sin\iota \cos\iota \cos{\left(\Phi(t)+ \Phi_{21}^C\right)} +  \\
& & F^D_\times(\alpha, \delta, \psi; t)\sin\iota  \sin{\left(\Phi(t)+ \Phi_{21}^C\right)} \Big] ,  \\
\nonumber
h_{22} &= - C_{22} \Big[&F^D_+(\alpha, \delta, \psi; t) (1 + \cos^2\iota) \cos{\left(2\Phi(t)+ \Phi_{22}^C\right)} +   \\
& & 2 F^D_\times(\alpha, \delta, \psi; t) \cos\iota  \sin{\left(2\Phi(t)+ \Phi_{22}^C\right)} \Big] .
\end{eqnarray}
In these equations, $C_{21}$ and $C_{22}$ are dimensionless constants that give the amplitudes of the components.  The angles $(\alpha, \delta)$ are the right ascension and declination of the source, while the angles $(\iota, \psi)$ specify the orientation of the star's spin axis relative to the observer.  The quantities $\Phi_{21}^C, \Phi_{22}^C$ are phase angles.  The functions $F^D_+$ and $F^D_\times$, known as the antenna or beam functions, describe how the two polarization components of the signal project onto the detector \citep[see, e.g.,][]{1998PhRvD..58f3001J}. The quantity $\Phi(t)$ is the rotational phase of the source.

The special and familiar case of single harmonic emission from a steadily spinning triaxial star is obtained by setting $C_{21} = 0$, leaving only the higher frequency component.  In this case, the amplitude is more conventionally parameterized as the dimensionless $h_0$, the amplitude of the (circularly polarized) signal that would be received if the star lay directly above or below the plane of the detector, with its spin axis pointing directly towards (or away from) the detector, so that $h_0 = 2C_{22}$.  Such triaxial stars are often colloquially described as having `mountains', or having a dimensionless equatorial ellipticity $\epsilon$ defined in terms of its principal moments of inertia $(I_{xx}, I_{yy}, I_{zz})$:
\begin{equation}
    \epsilon \equiv \frac{|I_{xx} - I_{yy}|}{I_{zz}} ,
\end{equation}
with the understanding that the star spins about the $z$-axis.  
The \gwh amplitudes and equatorial ellipticities are then related by
\begin{equation}\label{eq:h0}
    h_0 = \frac{16\pi^2 G}{c^4} \frac{I_{zz} \epsilon f_{\rm rot}^2 }{d} ,
\end{equation}
where $f_{\rm rot}$ is the rotational frequency and $d$ the star's distance.  Yet another quantity that is often quoted is the mass quadrupole $Q_{22}$, a quantity with the same dimension as the moment of inertia, and one which appears directly in the mass quadrupole formalism for calculating gravitational wave amplitudes:
\begin{equation}
    Q_{22} = I_{zz} \epsilon \sqrt{\frac{15}{8\pi}} .
    \label{eq:q22}
\end{equation}
When applying these formulae, we will use a fiducial value $I_{zz}^{\rm fid} = 10^{38}$\,kg\,m$^2$ for the moment of inertia.

We quote our results in terms of the ratio between minimum \gwh detectable amplitude and the spin-down limit, which is given by:

\begin{equation}
    h_{0,{\rm sd}}=\frac{1}{d}\left(\frac{5GI_{zz}}{2c^3}\frac{|\dot{f}_{\rm rot}|}{f_{\rm rot}}\right)^{1/2} ,
\end{equation}
which comes from the assumption that all rotational energy lost by the pulsar powers the gravitational wave emission. This limit is surpassed when the minimum detectable \gw amplitude $h_0$ is smaller than $h_{0,\rm sd}$.

We also make a distinction between intrinsic and observed spin-downs of the pulsars we analyze. The observed spin-downs are affected by the transverse velocity of the source \citep{shklovskii1970possible}, and can differ substantially from the intrinsic ones (see Table~\ref{tab:pulsars}). So when possible, we use the intrinsic spin-down to calculate the spin-down limit.

In the case of the narrowband search, a range of frequencies and spin-down rates is searched over, centered on the rotationally-derived values, allowing for fractional deviations of up to a maximum value.  For emission close to $2f_{\rm rot}$ this corresponds to ranges in search frequency $f_{\rm GW}$ and its first time derivative $\dot f_{\rm rot}$ of:
\begin{eqnarray}
1 - \delta < &\frac{f_{\rm GW}}{2f_{\rm rot}}& 
< 1 + \delta , \label{eq:nbd1} \\
1 - \delta < &\frac{\dot f_{\rm GW}}{2\dot f_{\rm rot}}& 
< 1 + \delta .
\label{eq:nbd2}
\end{eqnarray}
Previous narrowband searches used values of $\delta$ of the order $\sim \mathcal{O}(10^{-4})$ motivated partly by astrophysical considerations for the \gwh emission mechanism. In fact, Equations~\eqref{eq:nbd1} and \eqref{eq:nbd2} can take into account the possibility that the \gw is emitted by a free precessing bi-axial neutron star \citep{2002MNRAS.331..203J} or the possibility  that the star crust and core are linked by a torque that would enforce corotation. In the previous cases, the \gw emitted would be a nearly monochromatic signal emitted at a slightly different frequency and spin-down with respect to the one observed from electromagnetic observations.
Section \ref{sec:methods} below gives further details of how these signal models are used by the various data analysis methods.  We note that the values of $\delta$ chosen for the present search are sufficient to cover a parameter range roughly an order of magnitude greater than what is expected astrophysically by the above mechanisms.

\section{Analysis methods}\label{sec:methods}

Here, we briefly describe the analysis methods used in producing our results. We highlight any
differences in the methods compared to those used in previous analyses
\citep[e.g.][]{2019ApJ...879...10A,2019PhRvD..99l2002A}. For the analyses presented here, the
methods are variously applied for two different signal models: i) a signal emitted purely by the
$l=m=2$ mass quadrupole mode (i.e., a rigid triaxial rotator) at precisely, or close to, twice the
star's rotation frequency, and ii) a signal emitted by one or both of the $l=m=2$ and $l=2$, $m=1$
modes with components at precisely, or close to, once and twice the rotation frequency. For the
searches that do not allow a narrowband of frequencies and frequency derivatives, we assume that
the best fit radio timing model gives a phase coherent solution over the full range of the \gwh data
and we do not account for any uncertainties on the radio-derived values.

The methods for targeted searches assume that the \gwh signal precisely tracks the 
radio-derived phase evolution and therefore only a single phase evolution template is required. In the following sections we describe the three methodologies employed in this paper: \textit{The time-domain Bayesian method}, \textit{the $\mathcal{F}/\mathcal{G}$-statistic} method and the \textit{$5n$-vector method}.
The first two methods coherently analyze O1, O2 and O3 data\footnote{with the exception of the Vela pulsar.}, while the latter, along with the \textit{$5n$-vector narrowband} search, use only O3 data (see Section~\ref{sec:gwdata} for more details on GW data). 

The analyses also consider the occurrence of pulsar glitches using different methodologies.
For the Crab pulsar (J0534+2200), there were five glitches over the analysis period \citep[see
Section~\ref{sec:datayoung} and Section~2.1.1 of][]{2019ApJ...879...10A}; for the Vela pulsar, there was a glitch between O2 and O3a (\cite{gancio2020upgraded} and references therein).

\subsection{Time-domain Bayesian method}\label{sec:methodsbayesian}

As described in \citet{Dupuis:2005}, for each pulsar this method preprocesses the raw \gwh strain, which is then used as the input to a Bayesian parameter estimation code
\citep{2017arXiv170508978P}. The parameter estimation uses a nested sampling algorithm, as implemented in the {\sc
LALInference} package \citep{2010PhRvD..81f2003V, PhysRevD.91.042003}, 
to infer the unknown \gwh parameters of the expected signal, which depend on the signal model
described Section~\ref{sec:model}.
In contrast to the previous searches for the $l=m=2$ mode using this method
\citep[e.g.][]{2014ApJ...785..119A, 2017ApJ...839...12A, 2019ApJ...879...10A}, which have directly
inferred the \gwh amplitude $h_0$ for each signal, we now parameterize the amplitude in terms of the
mass quadrupole $Q_{22}$ and pulsar distance $d$ as in Equations~(\ref{eq:h0}) and (\ref{eq:q22}).
The distances are given Gaussian prior probability distributions, with mean and standard deviation
values taken from the distance estimates for the pulsars (see Table~\ref{tab:pulsars}). The $Q_{22}$
prior distribution is chosen to be flat over the range $[0, \scinum{5}{37}]$\,kg\,m$^2$, and zero
outside this range. This is not a physically motivated range, but is chosen to be more than an order
of magnitude larger than the largest upper limit found in \citep{2019ApJ...879...10A}.

In the \gwh analysis we
assume that the signal evolution is affected by a glitch in the same way as that observed with the
electromagnetic pulses, except that each glitch may introduce a phase offset between
the electromagnetic and \gwh signals. These unknown phase offset
parameters are included in the parameter inference. Three of the Crab pulsar glitches described in Section~\ref{sec:datayoung} occurred between O2 and O3, so it would be impossible to use our \gwh data to distinguish different phase offsets for each of these glitches. Therefore, only one phase offset parameter is required to account for the three glitches. During this work an error was found and fixed in the
analysis when accounting for the glitch behaviour during the parameter inference stage. This led to the time-domain Bayesian results for the Crab and Vela pulsar from \citet{2019ApJ...879...10A} being updated to those now given in \citet{erratum}.

As described in Section~\ref{sec:datayoung}, for the Vela pulsar we have a coherent timing
model over only the period of O3a. Therefore, we have to combine the results from an analysis on O1 and
O2 data with that from O3a in a semi-coherent manner. This also means that we do not need to account for the Vela pulsar glitch between O2 and O3a with the inclusion of an additional phase offset. Because of the bug described above, an analysis of
combined O1 and O2 data used in \citep{2019ApJ...879...10A} was repeated for this work, but with the
corrected code and (for the single harmonic search) with parameter inference on $Q_{22}$ and
distance instead of $h_0$. For the single harmonic search, the joint posterior on $Q_{22}$ and
$\iota$ was fitted with a multivariate Gaussian Mixture Model \citep[using the {\tt
BayesianGaussianMixture} function within scikit-learn][]{scikit-learn}, allowing a maximum of twenty
components. This mixture model was then used as the prior on these parameters when analysing O3a
data. For the dual harmonic search the mixture model was fitted to the joint $C_{21}$, $C_{22}$ and
$\iota$ posterior.

\subsection{Time-domain $\mathcal{F}$/$\mathcal{G}$-statistic method}\label{sec:methodsfstat}

The time-domain  $\mathcal{F}$/$\mathcal{G}$-statistic method uses the $\mathcal{F}$ and $\mathcal{G}$ statistics
developed in  \citet{1998PhRvD..58f3001J} and \citet{Jaranowski:2010}. 
 The $\mathcal{F}$-statistic is used when the amplitude, phase and polarization of the signal are unknown, whereas the $\mathcal{G}$-statistic is applied when only amplitude and phase are unknown, and the polarization of the signal is known (as described in Section~\ref{sec:restricted}). The methods have been used in several analyses of LIGO and Virgo data \citep{2011ApJ...737...93A, 2014ApJ...785..119A, 2017ApJ...839...12A}.

In this method a signal is detected in the data
if the value of the $\mathcal{F}$- or $\mathcal{G}$-statistic exceeds a certain threshold corresponding to an acceptable false alarm probability. We consider the false alarm probability of 1\% for the signal to be significant. 
The $\mathcal{F}$- and $\mathcal{G}$-statistics are computed for each detector and each inter-glitch period separately. The results from different detectors or  different inter-glitch periods are then combined incoherently by adding the respective statistics.
When the values of the statistics are not statistically
significant, we set upper limits on the amplitude of the
gravitational-wave signal.

\subsection{$5n$-vector method}\label{sec:methods5vector}

The frequency-domain $5n$-vector method has been introduced in \citet{2010CQGra..27s4016A, Astone:2012} and used in several analyses of LIGO and Virgo data \citep{2011ApJ...737...93A, 2014ApJ...785..119A, 2017ApJ...839...12A, 2019ApJ...879...10A}. It is also at the basis of the narrowband pipeline described in Section~\ref{sec:narrowband}. 
In this paper it has been applied to a subset of three pulsars: J0711\textminus6830, the Crab pulsar and the Vela pulsar.

In contrast to past analyses -- which used resampling -- the barycentric, spin-down and Einstein delay corrections are done by heterodyning the data, using the Band Sampled Data (BSD) framework \citep{2019CQGra..36a5008P}. This significantly reduces the computational cost of the analysis, which drops from about half of a CPU-day to a few CPU-minutes per source per detector. 
A detection statistic, based on the matched filter among the 5n-vectors of the data and the signal, is obtained and used to estimate the significance of an analysis result. Upper limits are computed using the approach first introduced in \citep{2014ApJ...785..119A}.

As in \cite{2019ApJ...879...10A}, two independent analyses have been done assuming the emission takes place at two times the star rotation frequency and at the rotation frequency \citep[according to the model described in][]{2010MNRAS.402.2503J}. 
While performing this analysis, we identified an incorrect choice for the range of amplitudes used to inject simulated signals in the O2 analysis of the pulsar J0711\textminus6830, see \citet{erratum} for more details.
This affects only the upper limit computation at the rotation frequency for J0711\textminus6830,  and the corrected value is given in \citet{erratum}. 

\subsection{Restricted orientations}\label{sec:restricted}

As with previous analyses, all of the pipelines produce results for the Crab and Vela pulsars based on
two different assumptions. The first is that the orientation of the pulsar is unknown, so a uniform
prior over the inclination and polarization angle space is used. The second uses estimates of the
source orientation based on X-ray observations of the pulsar wind nebulae tori \citep{Ng:2004es,
Ng:2008}, which are included in the pipelines as narrow priors on inclination and polarization
angle (effectively defining the polarization state of the signal), as given in Table~3 of \citet{2017ApJ...839...12A}.

\subsection{$5n$-vector narrowband}\label{sec:narrowband}

The {\it $5n$-vector narrowband pipeline} described in \citet{2017CQGra..34m5007M} uses the $5n$-vector method of \citet{2010CQGra..27s4016A,Astone:2012} and expands it to a narrow frequency and spin-down range around the source ephemerides values. 
This pipeline has previously been applied to the O1 and O2 datasets in \citet{2017PhRvD..96l2006A, O2Narrowband} permitting the analysis of pulsars for which ephemerides were not accurately known.

In contrast to \citet{O2Narrowband}, we now combine the matched filter's results between the detectors using weight factors computed from the power spectral density: each dataset is weighted inversely by the median noise power in the analyzed frequency band.  This allows the analysis to depend most strongly on the most sensitive dataset.
The final step is to select the local maximum of a detection statistic every $10^{-4}$\,Hz over the spin-down values considered.
Within this set of points in the parameter space, we select as outliers those with a $p$-value below a 0.1\% threshold (taking into account the number of trials).

This method targets pulsars J0711\textminus6830, Crab and Vela. 
For J0711\textminus6830 and Vela we analyzed 6 months of data, so the frequency and spin-down resolutions were $\scinum{6.5}{-8}\,\rm{Hz}$ and $\scinum{4.3}{-15}\,\rm{Hz}\,\rm{s}^{-1}$, respectively.
For Crab the resolutions were $\scinum{1.0}{-7}\,\rm{Hz}$ and $\scinum{1.1}{-14}\,\rm{Hz}\,\rm{s}^{-1}$ since we considered only data preceding the glitch ($\sim 115$ days). The narrowband resolutions relate to the natural discretization step of the discrete Fourier transform. The resolution ensures that a nearly monochromatic \gwh signal, emitted in the explored parameter space, is subject to a maximum loss of signal-to-noise ratio of $\sim 36\%$ \citep{2002AJ....124.1788R}. Note that, in order to reduce this loss, an half-bin interpolation of the Fourier transform is implemented in the code.

For each pulsar, we analyze a gravitational-wave frequency and spin-down range set to within 0.4\% of the ephemerides frequency and spin-down. This corresponds\footnote{Note that for the frequency range of J0711\textminus6830 we used a value of 0.2\% with a corresponding $\delta \sim \scinum{1}{-3}$.  This was due the constraints given by the 1\,Hz subsampling procedure.} to $\delta \sim \scinum{2}{-3}$ in Equations ~\eqref{eq:nbd1} and \eqref{eq:nbd2}. With respect to the O2 narrowband search, this corresponds to a volume explored in the frequency/spindown range of 4 times larger. 
We report the frequency and spin-down bands explored in Table~\ref{tab:nb_space}.

\begin{deluxetable}{lcccc}
    \tablehead{
        \colhead{Pulsar} & \colhead{$\Delta f_{\rm GW}$} & \colhead{$ \Delta  \dot{f}_{\rm GW}$} &  \colhead{$n_f$} & \colhead{$n_{\dot{f}}$} \\[-6pt]
        \colhead{~} & \colhead{(Hz)} & \colhead{(Hz\,s$^{-1}$)} & 
        \colhead{~} & 
        \colhead{~}
    }
    \tablecaption{Frequency/spin-down ranges explored in the $5n$-vector narrowband search. Second and third columns: frequency and spin-down ranges explored. Fourth and fifth columns: number of values in frequency and number of spin-down values considered. The total number of templates per pulsar is $n_f \times n_{\dot{f}}.$
     \label{tab:nb_space}
    }
    \startdata
    J0534+2200\tablenotemark{a}  (Crab) & 0.24 & \scinum{3.0}{-12} &\scinum{3.8}{6}  & 270 \\
    J0711\textminus6830 & 0.72 & \scinum{8.4}{-15} &\scinum{1.2}{7}  & 3 \\
    J0835\textminus4510 (Vela)  & 0.10 & \scinum{1.4}{-13} &\scinum{1.4}{6}  & 33 \\
    \enddata
    \tablenotetext{a}{Only data before the glitch reported in \citet{2019ATel12957....1S} are considered.}
\end{deluxetable}

Finally, for computing the 95\% confidence level upper limits on the \gwh amplitude $h_0$ we use the procedure described in \citet{O2Narrowband} to inject several simulated \gwh signals in each $10^{-4}$~Hz sub-band. For each sub-band we set the upper limit at the strain amplitude for which 95\% of the injected signals are recovered.

\section{Data sets used} \label{sec:data}

\subsection{Gravitational-wave data}\label{sec:gwdata}

We use a combination of data from the first, second and third observing runs
of the Advanced LIGO \citep{2015CQGra..32g4001L} and Virgo \citep{2015CQGra..32b4001A} gravitational wave
detectors. 
For O1 and O2 only data from the LIGO Hanford (H1) and LIGO Linvingston (L1) detectors have been used, while for O3 data from both LIGO detectors and the Virgo (V1)
detector have been used.  The O1 data cover the period from 2015 September 11 to 2016 January 19, with duty
factors of $\sim51\%$ and $\sim 60\%$ for L1 and H1, respectively. The O2 data cover the period from 2016
November 30 to 2017 August 25, with duty factors of $\sim 57\%$ and $\sim 59\%$ for L1 and H1, respectively (including commissioning breaks).
For O3, a period from 2019 April 1 to 2019 October 1
was designated O3a, prior to a one month commissioning break. O3a had duty factors of $\sim 76\%$, $\sim
71\%$ and $\sim 76\%$ for L1, H1 and V1, respectively. 

The data and subsequent upper limits are subject to uncertainty in the calibration of the instruments. The calibration uncertainty varies in amplitude and phase over the course of a run. We do not account for these variations in our results (see below), but we expect them to have a negligible impact on the results.
For more details of the O1 and O2 data and calibration used in these searches see the discussions in \citet{2017ApJ...839...12A} and \citet{2019ApJ...879...10A}. The full raw strain data from the O1 and O2 runs are publicly available from the Gravitational Wave Open Science Center\footnote{\url{https://www.gw-openscience.org/data}} \citep{2015JPhCS.610a2021V,2019arXiv191211716T}.
For the LIGO O3a data set, the time-domain Bayesian and $\mathcal{F}/\mathcal{G}$-statistic methods use the ``C01'' calibration for LIGO, while the $5n$-vector methods use the ``C00'' calibration. 
The C01 calibration has estimated maximum amplitude and phase uncertainties of 7~\% and 4~$\deg$ \citep{callib} while the C00 estimates are 8~\% and 5~$\deg$.
For the Virgo O3a data set, all of the pipelines use the ``V0'' calibration with estimated maximum amplitude and phase uncertainties of 5~\% and 3~$\deg$.

For the Bayesian analysis we estimate that the statistical uncertainty on the upper
limits due to the use of a finite number of posterior samples is on the order of 1\%. 
For the $5n$-vector analysis the statistical uncertainty on the upper limits has been estimated to be 1-3\% depending on the target.

\begin{figure}[h!]
    \centering
    \includegraphics[scale=0.55]{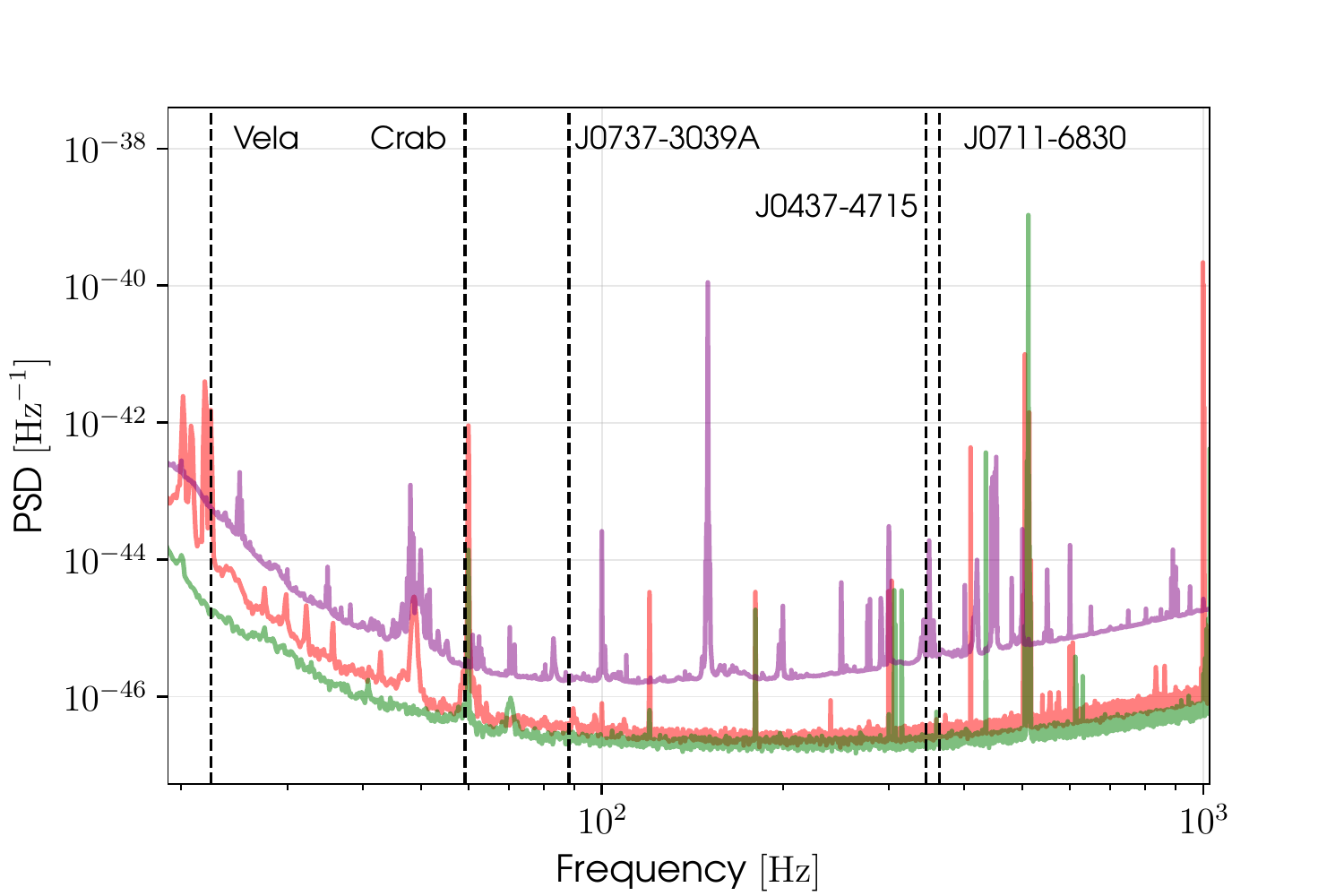}
    \caption{O3a noise PSD for H1, L1 and V1 in red, green and purple. H1 and L1 PSDs are calculated during a time period of optimal performance for the detector, while Virgo PSD is averaged over the run. The vertical dashed lines indicate the searched frequency region for each of the five pulsars.}
    \label{fig:spectra}
\end{figure}

Besides calibration uncertainties, the detectors' data sets are polluted by several noise disturbances.
Some of these disturbances are qualitatively visible as spikes or other deviations from smoothness in the noise power spectral densities (PSDs) for L1, H1 and V1 in Figure~\ref{fig:spectra} with respect to the five pulsars frequency searched.

\begin{deluxetable*}{lccccc}
    \tablehead{
        \colhead{Pulsar} & \colhead{$f_{\rm rot}$} & \colhead{$\dot{f}_{\rm rot}$} & \colhead{$\dot{f}_{\rm rot}^{\rm int}$} & \colhead{distance} & \colhead{Spin-down} \\[-6pt]
        \colhead{~} & \colhead{(Hz)} & \colhead{(Hz\,s$^{-1}$)} & \colhead{(Hz\,s$^{-1}$)} & \colhead{(kpc)} & \colhead{luminosity (W)}
    }
    \tablecaption{The properties of the pulsars in this search.
     \label{tab:pulsars}
    }
    \startdata
    \hline \\
\noalign{\vskip -2.5ex}
\multicolumn{6}{c}{Young pulsars} \\
\noalign{\vskip .8ex}
\hline \\
\noalign{\vskip -3ex}
    J0534+2200 (Crab) & 29.6 & \scinum{-3.7}{-10} & \nodata & $2.0\pm0.5$\tablenotemark{$^a$} & \scinum{4.5}{31} \\
    \noalign{\vskip .5ex}
    J0835\textminus4510 (Vela) & 11.2 & \scinum{-2.8}{-11}\tablenotemark{$^b$} & \nodata & $0.287_{-0.017}^{+0.019}$\tablenotemark{$^c$} & \scinum{6.9}{29} \\
    \hline \\
\noalign{\vskip -2.5ex}
\multicolumn{6}{c}{Recycled pulsars} \\
\noalign{\vskip .8ex}
\hline \\
\noalign{\vskip -3ex}
    J0437\textminus4715 & 173.7 & $\scinum{-1.7}{-15}$ & \scinum{-4.1}{-16} & $0.15679\pm0.00025$\tablenotemark{$^d$} & $\scinum{2.8}{26}$ \\
    \noalign{\vskip .5ex}
    J0711\textminus6830 & 182.1 & \scinum{-4.9}{-16} & \scinum{-4.7}{-16} & $0.110\pm0.044$\tablenotemark{$^e$} & \scinum{3.4}{26} \\
    \noalign{\vskip .5ex}
    J0737\textminus3039A & 44.1 & \scinum{-3.4}{-15} & \nodata & $1.15^{+0.22}_{-0.16}$\tablenotemark{$^f$} & \scinum{5.9}{26}
    \enddata
    \tablecomments{If an intrinsic rotation period derivative $\dot{P}_{\rm rot}^{\rm int}$ is available from the ATNF Pulsar Catalog \citep{2005AJ....129.1993M}, and is significantly different from the observed value, then this is converted into an intrinsic frequency derivative via $\dot{f}_{\rm rot}^{\rm int} = -f_{\rm rot}^2 \dot{P}_{\rm rot}^{\rm int}$ and is quoted here. For J0437\textminus4715 and J0711\textminus6830  this intrinsic frequency derivative will be used to calculate the spin-down luminosity and the spin-down limits in Table~\ref{tab:targetedresults}.}
    \tablenotetext{a}{\citet{2008ApJ...677.1201K}}
    \tablenotemark{b}{The $\dot{f}_{\rm rot}$ value given here is for the observation span used in this work, howevereb the spin-down limit shown in Table~\ref{tab:targetedresults} uses the long-term value of $f_{\rm rot} = -\scinum{1.57}{-11}$\,Hz\,s$^{-1}$ as given in the ATNF Pulsar Catalog \citep{2005AJ....129.1993M}.}
    \tablenotetext{c}{This distance is from \citet{2003ApJ...596.1137D}, although the Bayesian analysis described in Section~\ref{sec:methodsbayesian} uses a symmetric distance uncertainty of $0.288\pm0.018$\,kpc.}
    \tablenotetext{d}{\citet{2016MNRAS.455.1751R}}
    \tablenotetext{e}{This distance is based on dispersion measure from the \citet{2017ApJ...835...29Y} model, with a 40\% uncertainty assumed.}
    \tablenotetext{f}{This distance is from \citet{2009Sci...323.1327D}, although the Bayesian analysis described in Section~\ref{sec:methodsbayesian} uses a symmetric distance uncertainty of $1.18\pm0.19$\,kpc.}
\end{deluxetable*}

\subsection{Electromagnetic data}\label{sec:emdata}

The timing solutions used in our \gwh searches have been derived from electromagnetic observations of pulsars. These pulsars' basic properties are given in Table~\ref{tab:pulsars}, and are further explained in the next subsections.

\subsubsection{Recycled pulsars}

The pulsars J0437\textminus4715 and J0711\textminus6830 are monitored by the Parkes Pulsar Timing Array project \citep[PPTA;][]{Manchester+13}. The timing models for these pulsars were determined using data from the second data release of the PPTA \citep[DR2;][]{Kerr+20}. The model parameters were fit using {\sc Tempo2} \citep{2006MNRAS.369..655H}, with the stochastic red noise and dispersion measure (DM) variations characterised as power law processes and included in the fit (as the phases of a series of Fourier components for each power law). The power law parameters (amplitude and spectral index) and white noise properties were determined using the {\sc Enterprise}~\citep{Ellis+19} Bayesian pulsar timing analysis software. The noise models were consistent with those published with DR2. The timing stability for the pulsars J0437\textminus4715 and J0711\textminus6830 is such that the weighted root-mean-square (RMS) timing residual (excluding DM variations, but including spin noise) is 0.006\% and 0.035\% of a pulse period, respectively, over a span of $\sim$14\,years.


The timing model for the pulsar J0737\textminus3039A was developed using a combination of archival observations taken at various frequencies ranging between 604-1410\,MHz by the CSIRO 64-m Parkes radio telescope from 2004-2014, and 835\,MHz observations performed by the upgraded Molonglo Observatory Synthesis Telescope~\citep[UTMOST;][]{2017PASA...34...45B} between 2015 and 2018. TOAs at each observing band were computed via the standard cross-correlation technique, with each frequency band using its own template. They were then analyzed using the {\sc TempoNest}~\citep{2014MNRAS.437.3004L} Bayesian pulsar timing plugin to {\sc Tempo2}, which allowed us to measure the pulsar's deterministic and stochastic (red and white noise) properties simultaneously. The post-fit timing residuals have a weighted RMS of $\sim24\,\mu{\rm s}$, corresponding to about 0.01\% of a pulse period over $\sim$15\,years. 

\subsubsection{Young pulsars}\label{sec:datayoung}

As mentioned in Section~\ref{sec:methods}, the time domain Bayesian and $\mathcal{F}$/$\mathcal{G}$-statistic methods coherently analyze all O1, O2 and O3a data, while the $5n$-vector method only uses O3a data. Therefore, for the Crab pulsar, two timing solutions were obtained as described below: one using radio observations overlapping with O3a and another using data overlapping the period between the start of O1 and the end of O3a.

For the $5n$-vector search, the timing model for the Crab pulsar was created using pulse times-of-arrival (TOAs) measured at the Jodrell Bank Observatory (JBO) between April and October 2019. The dataset contains 352 TOAs obtained with the 42-ft telescope, using a 10\,MHz wide band, centred on 610\,MHz. In order to carefully track DM variations in the direction of the Crab pulsar, we include an additional 134 TOAs obtained with the 76-m Lovell telescope, using a 384\,MHz wide band, centred on 1520\,MHz. Further details of JBO observations can be found in \cite{2015MNRAS.446..857L}.

To account for the effects of timing noise on the Crab pulsar's rotation, we fit the TOAs, using {\sc Tempo2}, with a Taylor series of the spin frequency comprising terms up to 12th order. The Crab pulsar exhibits strong variations in DM, primarily due to the dynamics of the supernova remnant in which the pulsar resides \citep[e.g.,][]{2018MNRAS.479.4216M}. In order to mitigate the effects of DM variations on the measured TOAs from the Crab pulsar, we fit piece-wise the DM at 22 epochs within the O3a period, meaning the value of DM in the timing model is updated every $\sim$8 days.  Finally, we include in the timing model the effects of a moderately sized spin-up glitch which occurred during an observation of the Crab pulsar on 2019 July 23 \citep{2019ATel12957....1S}.  Applying this timing model to the measured TOAs, the resulting timing residuals have a RMS value of $\sim 67\,\mu{\rm s}$, corresponding to 0.2\% of one pulse period.

The second timing model for the Crab pulsar, used for the time domain Bayesian and $\mathcal{F}$/$\mathcal{G}$-statistic searches, was created covering the entire period from August 2015 to October 2019. In this case, the dataset comprises 2478 TOAs measured with the 42-ft telescope and 858 TOAs measured with the Lovell telescopes at the same bandwidths and centre frequencies as stated above, forming a total of 3336 observations. For these data, the timing noise was modelled using a Taylor series of the spin-frequency with terms up to 12th order, in combination with 100 harmonically related sinusoids, implemented using the {\sc FITWAVES} functionality in {\sc Tempo2}.  A piece-wise model of the DM was also included, comprising DM values at 110 epochs (approximately every 14 days). Over this time period, the Crab pulsar underwent five spin-up glitches including the July 2019 glitch and the largest glitch observed to date in the Crab pulsar, which occurred in November 2017 \citep{2018MNRAS.478.3832S}. These two glitches and their recoveries are included in the timing model. The remaining three glitches were sufficiently small as to be fully described by the other parameters together with the timing noise and so are not specifically modelled here. The residuals resulting from this timing model have an RMS value of $\sim$ 21 $\mu$s, corresponding to 0.06\% of one pulse period. 

A timing model for the Vela pulsar was created using pulse TOAs from the Mt Pleasant 26-m radio observatory near Hobart, Tasmania. The entire O3a observing period was covered and the centre frequency was 1376\,MHz with a bandwidth of 64\,MHz. The single-pulse observations were integrated to 1\,hr and {\sc Tempo2} was used to create an ephemeris from those 464 TOAs. A Taylor series to the 4th derivative was used to get an RMS of $\sim$50\,$\mu$s, which is 0.06\% of the pulse period.


\section{Analysis Results} \label{sec:results}

\subsection{Targeted searches}

The results from the targeted searches for all five pulsars are summarized in
Table~\ref{tab:targetedresults} with the three different pipelines  presented together for ease
of comparison. 

\startlongtable
\begin{deluxetable*}{lclcccccc}
\tablecaption{Limits on Gravitational-wave Amplitude, and Other Derived Quantities, for the Three Targeted Searches.\label{tab:targetedresults}}
\tabletypesize{\scriptsize}
\tablehead{
\colhead{Pulsar Name} & 
\colhead{$h_0^{\text{sd}}$} & 
\colhead{Analysis} & 
\colhead{$C_{21}^{95\%}$} & 
\colhead{$C_{22}^{95\%}$} & 
\colhead{$h_0^{95\%}$} & 
\colhead{$Q_{22}^{95\%}$} & 
\colhead{$\epsilon^{95\%}$} & 
\colhead{$h_0^{95\%}/h_0^{\text{sd}}$}\\[-6pt]
\colhead{(J2000)} & 
\colhead{($10^{-26}$)} & 
\colhead{Method} & 
\colhead{($10^{-26}$)} & 
\colhead{($10^{-27}$)} & 
\colhead{($10^{-26}$)} & 
\colhead{($10^{32}$\,kg\,m$^2$)} & 
\colhead{~} & 
\colhead{~}}
\startdata
\hline \\
\noalign{\vskip -2.5ex}
\multicolumn{9}{c}{Young pulsars\tablenotemark{$^a$}} \\
\noalign{\vskip .8ex}
\hline \\
\noalign{\vskip -3ex}
\multirow{3}{*}{J0534+2200\tablenotemark{$^b$} (Crab)} & \multirow{3}{*}{140} & Bayesian & 12.7(7.9) & 6.3(5.6) & 1.5(1.2) & 6.6(5.7) & \scinum{8.6(7.4)}{-6} & 0.010(0.009) \\ 
 &  & $\mathcal{F}$/$\mathcal{G}$-statistic & 8.9(6.2) & 7.9(7.1) & 1.9(1.5) & 7.9(6.3) & \scinum{10(8.1)}{-6} & 0.014(0.011) \\ 
 &  & 5$n$-vector & 15.9(12.4) & \nodata & 3.0(2.9) & 12.6(12.1) & \scinum{16.3(15.7)}{-6} & 0.021(0.021) \\
 \noalign{\vskip 2ex}
\multirow{3}{*}{J0835\textminus4510 (Vela)} & \multirow{3}{*}{330} & Bayesian & 1100(980) & 120(84) & 22(17) & 91(73) & \scinum{12.0(9.5)}{-5} & 0.067(0.052) \\ 
 &  & $\mathcal{F}$/$\mathcal{G}$-statistic & 1470(1370) & 116(48)& 23(12) & 96(50) & \scinum{12.4(6.4)}{-5} & 0.070(0.036) \\ 
 &  & 5$n$-vector & 1700(1400) & \nodata & 24(24) & 100(102) & \scinum{13.0(13.2)}{-5} & 0.073(0.073) \\
\hline \\
\noalign{\vskip -2.5ex}
\multicolumn{9}{c}{Recycled pulsars} \\
\noalign{\vskip .8ex}
\hline \\
\noalign{\vskip -3ex}
\multirow{3}{*}{J0437\textminus4715} & \multirow{3}{*}{0.79} & Bayesian & 2.2 & 4.1 & 0.78 & 0.0074 & \scinum{9.5}{-9} & 0.99 \\ 
 &  & $\mathcal{F}$/$\mathcal{G}$-statistic & 2.1 & 7.2 & 0.86 & 0.0082 & \scinum{11.0}{-9} & 1.1 \\ 
 &  & 5$n$-vector & \nodata & \nodata & \nodata & \nodata & \nodata & \nodata \\
 \noalign{\vskip 2ex}
\multirow{3}{*}{J0711\textminus6830} & \multirow{3}{*}{1.2} & Bayesian & 2.6 & 3.5 & 0.82 & 0.0064 & \scinum{8.3}{-9} & 0.68 \\ 
 &  & $\mathcal{F}$/$\mathcal{G}$-statistic & 2.4 & 9.4 & 0.98 & 0.0059 & \scinum{7.7}{-9} & 0.82 \\ 
 &  & 5$n$-vector & 2.9 & \nodata & 0.91 & 0.0053 & \scinum{7.2}{-9} & 0.76 \\
 \noalign{\vskip 2ex}
\multirow{3}{*}{J0737\textminus3039A} & \multirow{3}{*}{0.62} & Bayesian & 5.9 & 3.3 & 0.69 & 0.80 & \scinum{1.0}{-6} & 1.1 \\ 
 &  & $\mathcal{F}$/$\mathcal{G}$-statistic & 3.0 & 1.2 & 0.99 & 1.10 & \scinum{1.4}{-6} & 1.6 \\ 
 &  & 5$n$-vector & \nodata & \nodata & \nodata & \nodata & \nodata & \nodata \\
\enddata
\tablecomments{Parameters for the pulsars can be found in Table~\ref{tab:pulsars}.}
\tablenotetext{a}{For J0534+2200 and J0835\textminus4510 the results in parentheses are those when using restricted priors on the pulsar orientation.}
\tablenotetext{b}{For the $5n$-vector results only data from the O3a run was used for all the three pulsars.}
\end{deluxetable*}

\begin{figure}
\plotone{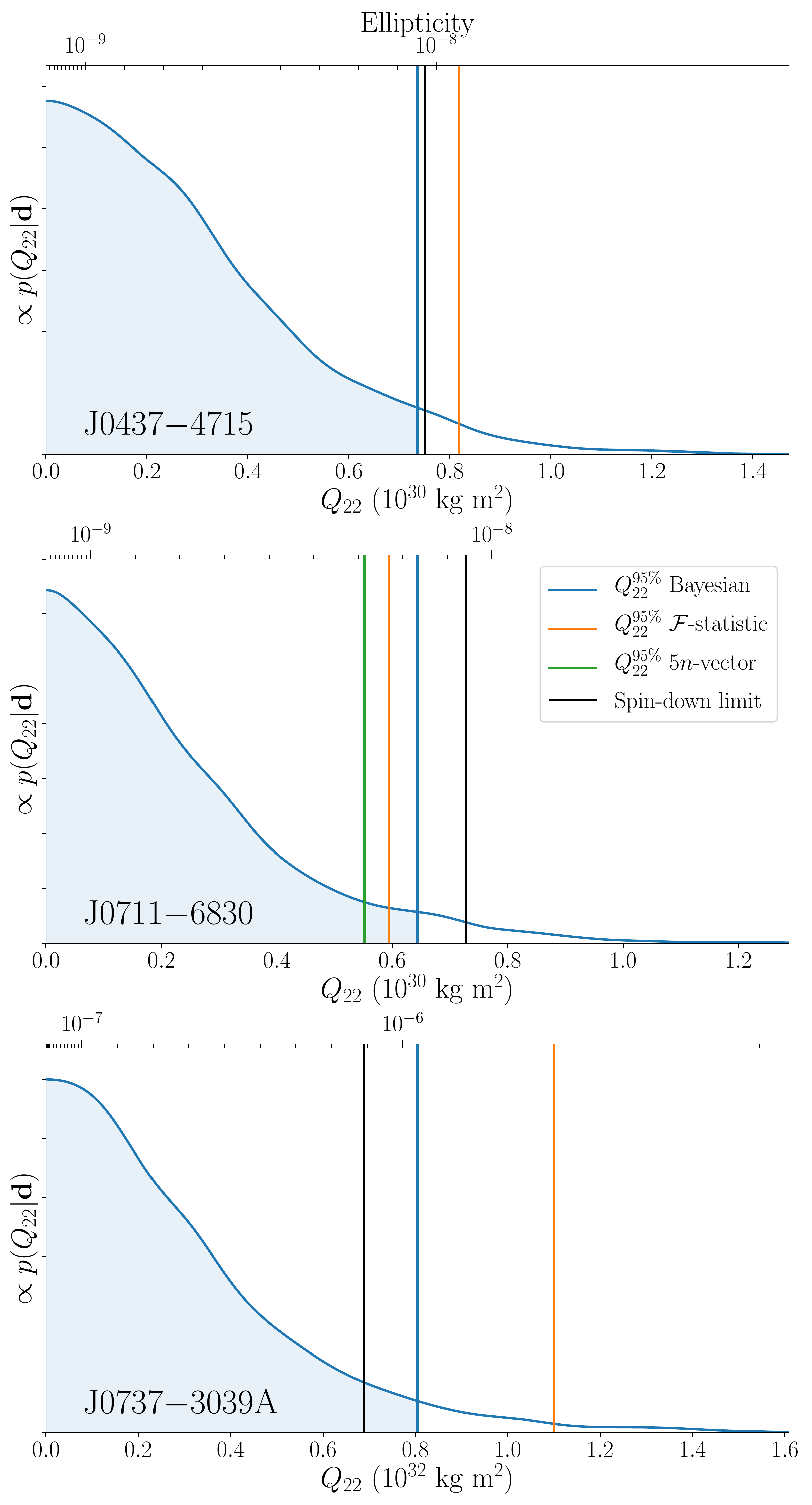}
  \caption{Posterior distributions on $Q_{22}$ for the Bayesian analysis of the three recycled pulsars. Also shown as vertical lines are the 95\% credible upper limits from the three different pipelines and the spin-down limits. The upper axes show the equivalent limits on the fiducial ellipticity.}
\label{fig:msps}
\end{figure}

No evidence for a signal was observed from any of the 5 pulsars in either of the
harmonics studied, so as with previous analyses we present the results as 95\% credible upper limits
on the \gwh amplitudes: $C_{21}$ and $C_{22}$ for the dual harmonic search, and $h_0$ for the single
harmonic search. For the single harmonic search, using Equations~(\ref{eq:h0}) and (\ref{eq:q22}),
we place equivalent limits on the mass quadrupole $Q_{22}$ and fiducial ellipticity $\epsilon$. The posterior distributions on $Q_{22}$ for the Bayesian analysis, along with
the point estimate upper limits from the $\mathcal{F}/\mathcal{G}$-statistic and $5n$-vector
pipelines, are shown in Figure~\ref{fig:msps} for the three recycled pulsars and Figure~\ref{fig:young} for Crab and Vela. In Figure ~\ref{fig:young} the upper limits are shown using both restricted-orientation and unrestricted priors as described in Section ~\ref{sec:restricted}.

Despite the analysis pipelines being largely independent, and the statistical procedures used to
derive upper limits being different, there is a broad agreement among the different pipelines. One
source of differences, however, comes from the pipelines not all using the same datasets. The $5n$-vector
search analyzed only O3a data while the Bayesian and $\mathcal{F}/\mathcal{G}$-statistic search
coherently (or semi-coherently in the case of the Vela pulsar) combined data from O1, O2 and O3a.
The methods of combining data for the Vela pulsar analysis are discussed in
Section~\ref{sec:methods}. 

Another source of differences is that the Bayesian analysis does not
assume a fixed distance, but instead includes it as a parameter to be estimated from the data.
Therefore, limits on $Q_{22}$ and $h_0$ are computed marginalizing over the distance, rather
than assuming a fixed value. In general, the distance posterior distributions match their priors
well but with a small bias towards larger values when the distance priors are wide. However, for J0711\textminus6830 the bias is more obvious with the peak in the distance
posterior being at a value approximately 20\% larger than that of the prior. This biasing of the distance is due to our choice of a flat prior on $Q_{22}$, which is not an uninformative distribution, i.e., there is much more prior weight for large $Q_{22}$ values, disfavoring smaller distances. 

In the discussions below we will generally refer to the most stringent, i.e., lowest, limit from the different searches and
will discuss only the single harmonic ($l=m=2$ mass quadrupole) and unrestricted-orientation priors results in detail.

\subsubsection{Recycled  pulsars}
We surpassed for the first time the spin-down limit for
J0437\textminus4715 and J0711\textminus6830. 
For J0437\textminus4715 our 95\% upper limits are just
below the spin-down limit and for J0711\textminus6830 it is at $\sim 70\%$ of the spin-down limit.
For J0711\textminus6830 this equates to less than half of the intrinsic spin-down energy loss being attributable
to \gwh emission. For these two pulsars, these limits provide constraints that
are below the fiducial ellipticity of $10^{-8}$, which was previously surpassed for only J0636+5129
\citep{2019ApJ...879...10A}. 

\subsubsection{J0737\textminus3039A}
For J0737\textminus3039A our limits are only just above the spin-down limit, and
would easily surpass it assuming a slightly larger moment of inertia. For this pulsar, despite
having a similar spin-down limit to two of the recycled pulsars, its significantly lower frequency and larger
distance leads to a limit on fiducial ellipticity that is around $10^{-6}$.

\subsubsection{Young pulsars \label{sec:young}}

The inclusion of O3a data for the Crab and Vela pulsars does not significantly improve the results
compared to previous analyses \citep{2019ApJ...879...10A, erratum}, because the detector sensitivity improvements achieved for the O3 run via quantum squeezing were greatest at high frequencies \citep{2019PhRvL.123w1107T, 2019PhRvL.123w1108A}.

\begin{figure}
  \plotone{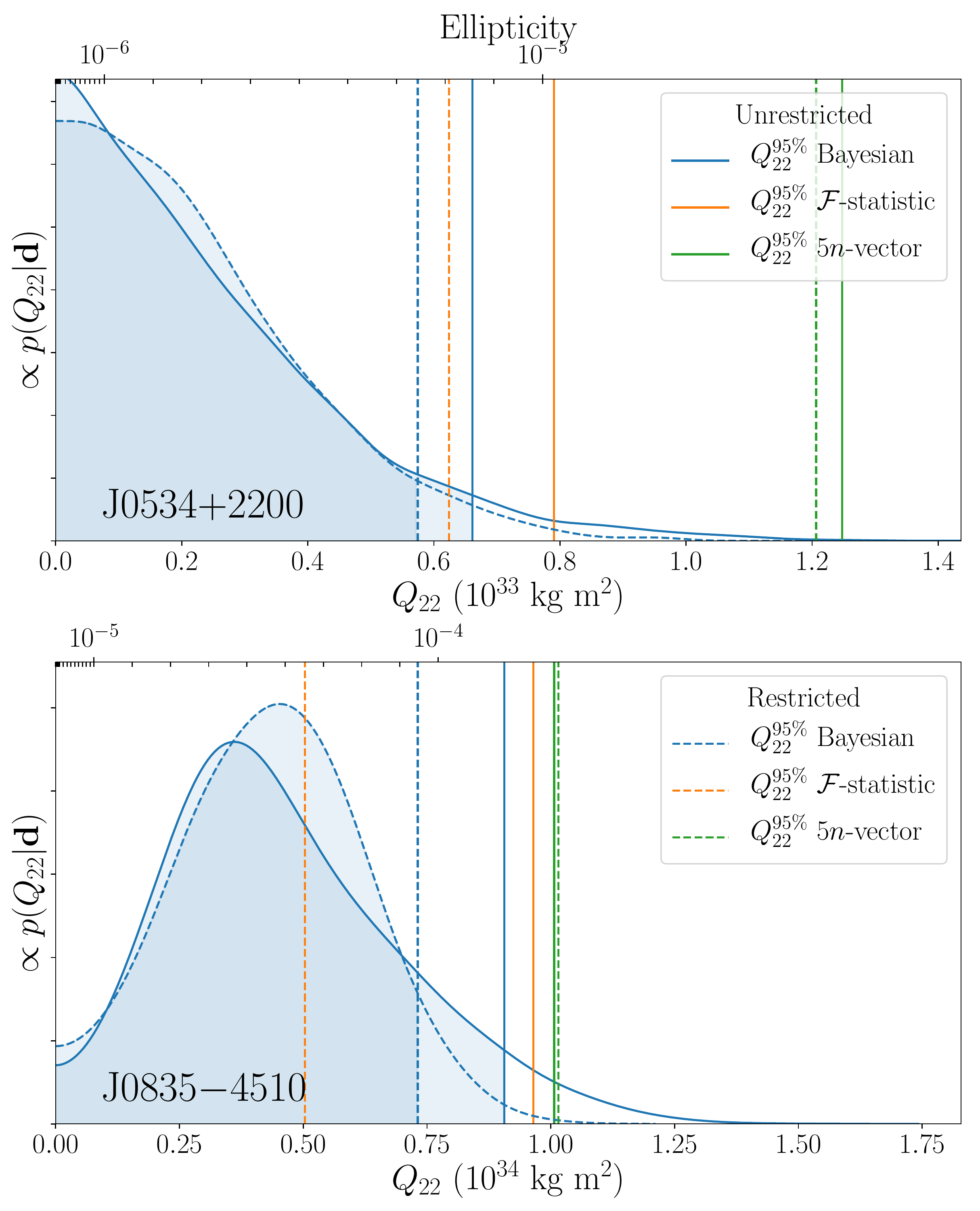}
  \caption{Posterior distributions on $Q_{22}$ for the Bayesian analysis of the two young pulsars Crab and Vela. The solid lines show the results when not using prior restrictions on the pulsar orientations (see Section~\ref{sec:restricted}), while those with restrictions are shown as the dashed lines. Also shown as vertical lines are the 95\% credible upper limits from the three different pipelines. The upper axes show the equivalent limits on the fiducial ellipticity.}
  \label{fig:young}
\end{figure}

We obtain limits on the GW emission from the Crab pulsar at 1-2\% of the spin-down limit regardless of prior choice, corresponding to a limit on fiducial
ellipticity of $\epsilon\sim 10^{-5}$.
For the Vela pulsar we obtain limits on the GW emission that are less than $\sim $7\% of the spin-down limit, corresponding to a fiducial ellipticity of $\epsilon \lesssim 10^{-4}$. 
As seen in Figure~\ref{fig:young}, the
posterior distribution on $Q_{22}$ peaks away from but is not disjoint from zero. Such a
distribution is not uncommon for pure Gaussian noise, but could also be due in part to spectral
contamination observed near twice the Vela pulsar's rotation frequency in all detectors (see Figures~\ref{fig:spectra} and \ref{fig:spectra_O3a_vela}).

\subsection{$5n$-vector Narrowband}

The narrowband search found no evidence for GW emission from J0711\textminus6830, the Crab pulsar or the Vela pulsar, although several analyses outliers were found for two of these pulsars.

\begin{figure}
    \centering
    \includegraphics[scale=0.36]{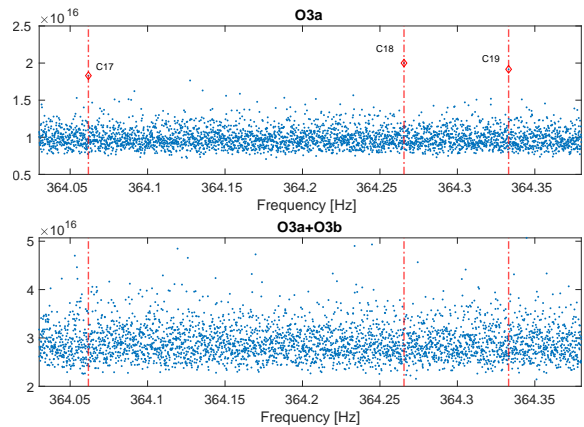}
    \caption{Top panel: Detection statistic obtained from the narrowband analysis of J0711\textminus6830 using O3a data. The outliers are indicated with red diamonds and red vertical lines. Bottom panel: Detection statistic obtained from the narrowband analysis for J0711\textminus6830 using the full O3 dataset. The frequencies of the original O3a outliers are indicated by red vertical lines.}
    \label{fig:ds}
\end{figure}

For J0711\textminus6830 there were 19 outliers. 
Sixteen outliers clustered at the boundaries of the analyzed frequency  band and were due to artifacts created by the band extraction close to the integer frequency of $364$\,Hz. 
These artifacts are created due to sub-sample processes at $1$\,Hz.
The remaining three outliers, labelled as C17, C18 and C19 (see Fig.~\ref{fig:ds} top panel), were found by the narrowband pipeline with a $p$-value of $ \sim  \scinum{1.2}{-7}$, which when rescaled for the number of trials corresponds to a $p$-value of $ \sim \scinum{5.0}{-4}$. 
In order to assess the significance of the outliers, we performed a narrowband search for  J0711\textminus6830 using the same setup for the rotational parameters but using an ``off-source'' sky-position. This procedure would effectively blind the analysis to the presence of a possible astrophysical signal, thus allowing us to build an empirical noise-only distribution of the detection statistic which can be used to reassess the outliers' significances.
From this analysis we found that the $p$-value of two of the three outliers were above the narrowband ceiling of 0.1\% (C17 and C19), while the value for C18 was 0.06\%. This test indicated that by re-assigning the significance with the off-source method, two of the outliers would not have passed the narrowband threshold for candidates selection and hence could be due to low-level noise instrumental artifacts.

The three outliers were followed-up by two of the targeted pipelines.
For the three outliers the time-domain Bayesian pipeline found a strong preference for the hypothesis that the data (with a bandwidth of $1/60$\,Hz centered on the outlier) was consistent with Gaussian noise compared to a hypothesis that it also contained signals coherent between detectors.
Specifically, the Bayes factors recovered for the signal vs.\ noise hypothesis were $<10^{-4}$, thus hinting for the noise origin for all of these outliers. 
Additionally, for C17 and C19, the pipeline recovered a maximum posterior on $h_0$ of $\scinum{1.6}{-26}$ and $\scinum{8.5}{-27}$. As argued in the case of the Vela pulsar in Sec.~\ref{sec:young}, this can be related to the presence of instrumental noise contributions.

The $5n$-vector targeted pipeline also performed a follow-up of the most significant of the three outliers (C18), with frequency $\sim 364.25$ Hz, using software injections with an amplitude set to that of one of the recovered outliers.
The pipeline found that the distribution of the software injection's detection statistic was compatible with the value displayed by the outlier. More precisely, for a set of 50 injected signals it found an average critical ratio $\overline{CR}=7.0$ with a standard deviation of 3.2, to be compared with a value $8.5$ found for the actual analysis candidate. In absence of any signal, the noise average critical ratio was found to be $\overline{CR}=0.3 \pm 1.3$.

Given that the previous tests did not conclusively establish the noise origin of the outliers, we performed a narrowband analysis using the full O3 LIGO dataset (C00 calibration). If the outliers were due to a real continuous wave signal, we would have expected to see them as more significant in this analysis.
Fig.~\ref{fig:ds} compares the detection statistics obtained from the narrowband analysis using only O3a data and the full O3 dataset. We see that the outliers found in the O3a run are no longer present when using the full run, inconsistent with an astrophysical signal.

\begin{figure}
    \centering
    \includegraphics[scale=0.5]{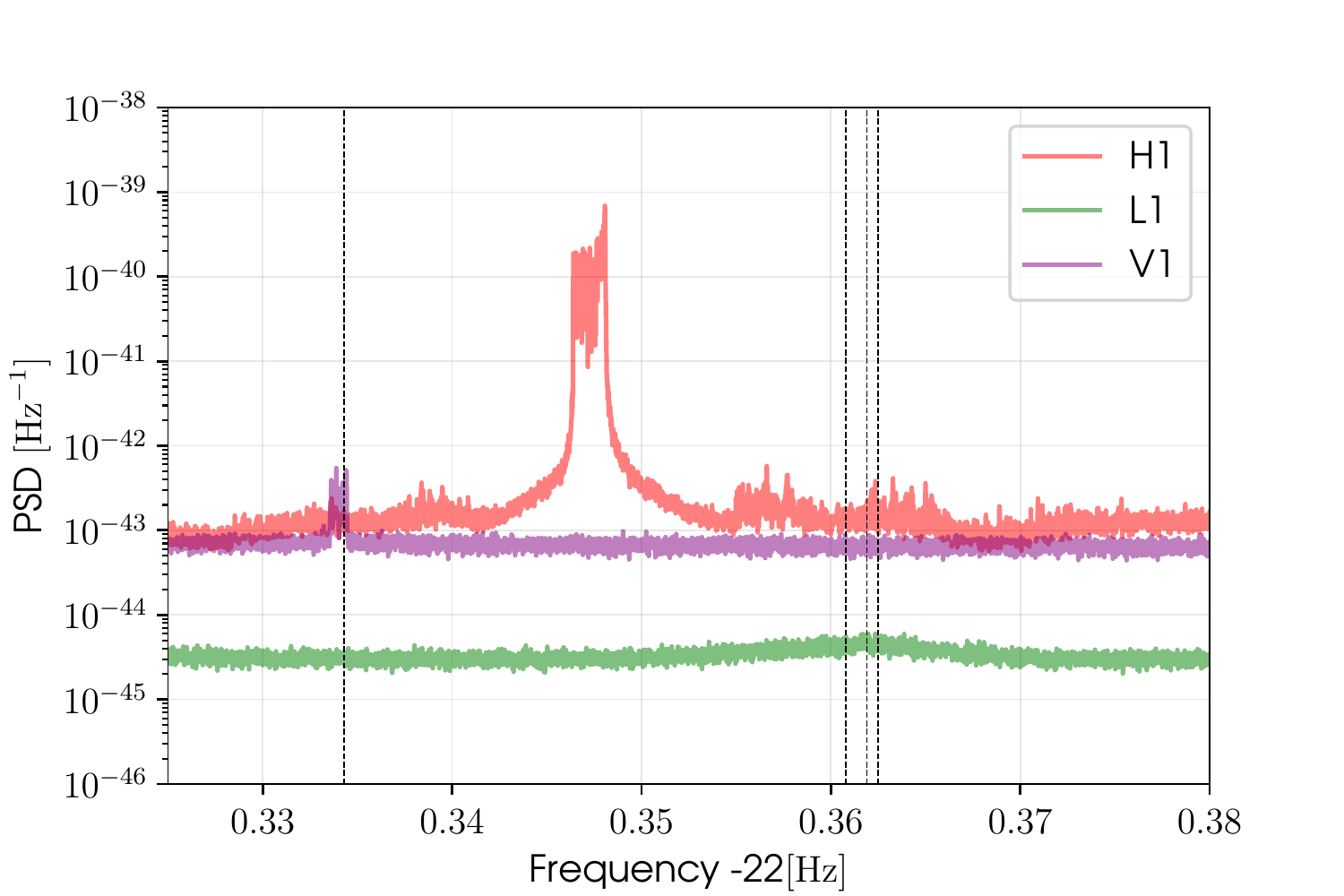}
    \caption{Power spectral density of H1 (red), L1 (green), V1 (purple) after the correction for the CW modulations. The dashed vertical lines mark the frequencies of the 4 Vela outliers affected by instrumental disturbances.}
    \label{fig:spectra_O3a_vela}
\end{figure}

Finally, for the Vela pulsar, we found four outliers, but these are due to noise disturbances in the data, see Figure~\ref{fig:spectra_O3a_vela}. One of the candidates was due to the left sidebands of a known H1 disturbance at 22.347 Hz together with a noise disturbance known V1 at 22.333 Hz. The others three outliers were due to the right sidebands of the H1 disturbance and a L1 broadband noise disturbance around 22.365 Hz.

\begin{figure}[h!]
\hspace{-.5cm}
\includegraphics[scale=0.6]{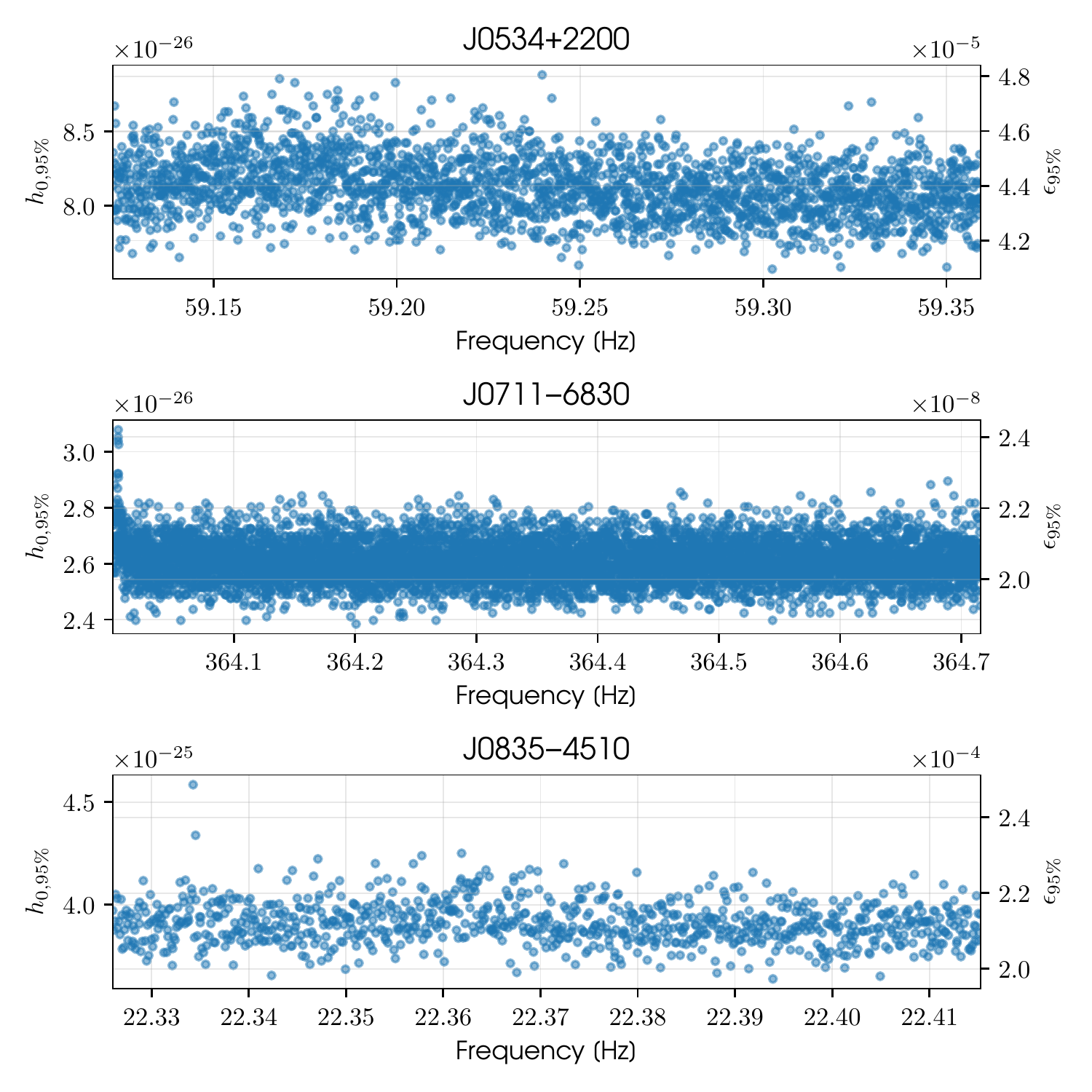}
\caption{The graphs show the 95\% confidence level upper limits on the \gwh amplitude $h_0$ and ellipticity $\epsilon$  for the three pulsars analyzed in the narrowband analysis. From top to bottom, the upper limits are shown for Crab, J0711\textminus6830  and Vela. The contribution of the H1 and V1 noise disturbances clearly visible in Vela's upper limits.}
\label{fig:NBUL}
\end{figure}

Hence we computed the 95\% confidence level upper limits on the \gwh amplitudes $h_0$ and the corresponding limits on the fiducial ellipticities, as shown in Figure~\ref{fig:NBUL}.
For pulsar J0711\textminus6830 we obtain median values of the upper limit on the amplitude $h_0$ and the ellipticity over the analyzed frequency band  of $\scinum{2.6}{-26}$ and $\scinum{2.0}{-8}$. Unfortunately, the narrowband pipeline does not surpass the spin-down limit for this pulsar.

For the Crab and Vela pulsars we obtain median values for the upper limit on $h_0$ of $\scinum{8.1}{-24}$ and $\scinum{3.90}{-25}$, while the corresponding median upper limits on the fiducial ellipticities are $\scinum{4.4}{-5}$ and  $\scinum{2.0}{-4}$.
These upper limits are a factor 1.6 and 2.25 better than the upper limit obtained with O2 data in \cite{O2Narrowband}.
This improvement has been partially possible by the inclusion of Virgo data and the slight improved sensitivities in O3 for the two LIGO detectors. Another major contribution, however, comes from the new PSD-weighted analysis that can account for data having different PSDs. 

\section{Discussion} \label{sec:discussion}

For the first time, we have been able to surpass the spin-down limit of a recycled pulsar. This achievement is significant for \gwh searches from known pulsars for two reasons.  First, the upper limits we have set on the ellipticities of these (rapidly rotating) stars are very small (around $10^{-8}$), a consequence of the scaling of wave amplitude with ellipticity and spin frequency, $h_0 \sim \epsilon f^2$.   Second and more crucial, recycled pulsars have a quite different evolutionary history from younger, more slowly spinning pulsars, as they are believed to have acquired their high spin frequencies in a prolonged period of accretion from a binary companion.
This sustained accretion can lead, in principle, to non-axisymmetric deformation of the star.

Several such accretion-specific deformation mechanisms are known.  One possibility was first noted by \citet{1998ApJ...501L..89B}, who argued that temperature asymmetries in the crust of an accreting star would produce lateral variations in the locations of the transition layers between one nuclear species and the next, a suggestion that has been examined in more detail since \citep{2000MNRAS.319..902U, 2020MNRAS.493.3866S, 2020MNRAS.tmp.1067O}.  Another possibility is that the accretion process `buries' the star's magnetic field, so that a very strong \emph{internal} field, much larger than the external field of $\sim 10^9$\,G inferred from a typical recycled pulsar, distorts the star \citep{2009MNRAS.395.1963V}.  Alternatively, it has been proposed that the changing shape of a centrifugally-distorted star could cause the crust to crack, either during the initial spin-up phase \citep{2018arXiv180404952F}, or during the later (post-accretion) spin-down phase \citep{1971AnPhy..66..816B}. The ellipticity would be generated if this cracking were to occur in a sufficiently non-axisymmetric way.  As a caveat, it should be noted that recycled pulsars are believed to be old, with ages $\sim 10^9$ years, giving much time for annealing of non-axisymmetric distortions.

One can convert our upper limits on ellipticity into approximate upper limits on the strain in the crust, or, alternatively, on the strength of the internal magnetic field, with the understanding that the limit applies only to the part of the strain or magnetic field that sources a \emph{quadrupolar} deformation of the star. 

Assuming crustal strain $u$, using Eq.~\eqref{eq:q22} of \citet{2000MNRAS.319..902U}, we have
\begin{equation}
\epsilon \approx 10^{-7} \frac{u}{10^{-2}} .
\end{equation}
To give two specific examples, this corresponds to a best  upper limit on the strain in the crust of the Crab pulsar of $u \approx 0.86$ (using the non-restricted priors), while for J0711\textminus6830, we have a best upper limit of $u \approx 7.2 \times 10^{-4}$.  This is to be compared with estimates of the breaking strain as high as $0.1$ from the molecular dynamics simulations of \citet{2009PhRvL.102s1102H} \citep[see also the semi-analytical calculations of][]{2018MNRAS.480.5511B}.  It should be noted however that application of such results to a real neutron star requires extrapolation through many orders of magnitude in both size and temporal duration as compared to the molecular dynamics simulations, so caution should be exercised.

This strain upper limit also underlines the significance of our new results for the recycled pulsars. While the spin-down limit for the Crab pulsar was surpassed some years ago \citep{2008ApJ...683L..45A}, the necessary crustal strain level required for a detection remains implausibly high, as noted above.  In contrast, only a small crustal strain would have been required to have produced a detectable level of \gwh emission from the recycled pulsars, which, together with surpassing the spin-down limit for our recycled pulsars, indicates that we are entering new territory in terms of the physical requirements for a detection.

Assuming instead distortion by a strong internal magnetic field $B_{\rm int}$ in a superconducting core, we can use the results of \citet{2012MNRAS.419..732L,2012MNRAS.421..760M, 2014MNRAS.437..424L}:
\begin{equation}
\epsilon \approx 10^{-8} \frac{B_{\rm int}}{10^{12} \, \rm G} .
\end{equation}
See also \citet{de_Araujo_2016,2017EPJC...77..350D} for another study related to distortion by a strong internal magnetic field for pulsars with known braking indices.
For the Crab pulsar this corresponds to a (non-restricted prior) upper limit on the internal field of $B_{\rm int} \approx \scinum{8.6}{14}$\,G.  This can be compared with the values of the \emph{external field}, as estimated assuming $100\%$ electromagnetic dipole spin-down, of $B_{\rm ext} \approx \scinum{3.8}{12}$\,G, as taken from the ATNF Pulsar Catalog \citep{2005AJ....129.1993M}, i.e., we can say that the internal magnetic field is no more than about $230$ times stronger than the inferred external field.  Similarly, for J0711\textminus6830 we have an upper limit $B_{\rm int} \approx \scinum{7.2}{11}$\,G, to be compared with the inferred $B_{\rm ext} \approx \scinum{2.9}{8}$\,G, i.e. we can say the internal field is no more than about $2,500$ times stronger than the inferred external field.  As noted above, a significantly larger internal field than external is possible, if field burial takes place during a previous accretion phase.  This would also require that the field is stable and remains buried over the lifetime of the star; see \citet{2017JApA...38...48M} for  a recent review of relevant issues for MSPs.

As in previous analyses \cite{2017ApJ...839...12A, 2019ApJ...879...10A} we have constrained the \gwh emission from the Crab pulsar.  For the Crab pulsar, we have set the upper limit on its ellipticity of $\approx \scinum{8.6}{-6}$.  While significantly larger than the ellipticity upper limits we have set for the recycled pulsars, this is nevertheless of considerable interest, as it represents an ellipticity of approximately  $\scinum{1.0}{-2}$ times the spin-down limit.  Equivalently, we can say that our non-detection implies that a fraction of no more than $\scinum{1.0}{-4}$ of the Crab pulsar's spin-down energy is going into the \gwh channel.  For the Vela pulsar, we have set a best upper limit of $\scinum{1.2}{-4}$ on its ellipticity, which is $\scinum{6.6}{-2}$ times its spin-down limit, showing than no more than $\scinum{4.4}{-3}$ of its spin-down energy is going into the \gwh channel.  Clearly, on energetic grounds, there was ample scope for making a detection, even if the required ellipticities themselves were comparatively large. 

The other results presented, including those of the narrowband search, give slightly less constraining upper limits on the \gwh amplitudes, with corresponding small changes in the inferred upper limits on ellipticity and fraction of energy going into the \gwh channel.


\section{Conclusions} \label{sec:conclusions}

In this paper, we have presented two main results.  We have reported new \gwh upper limits on the gravitational wave emission from the \acp{MSP}s J0437\textminus4715 and J0711\textminus6830, matching or surpassing their spin-down limits.  These limits represent a significant milestone for \gwh astronomy, as this is the first time our direct \gwh observations provide limits at or below the spin-down limit for a \MSP. We have also reported updated limits on the fraction of spin-down energy going into the gravitational wave channel for the two young pulsars, the Crab and Vela pulsars.

Recently, \citet{2018ApJ...863L..40W} noted a lack of pulsars at the bottom left of the pulsar period-period derivative diagram, i.e., a deficit in pulsars with high spin frequencies and small spin-down rates.  \citet{2018ApJ...863L..40W} noted this could be a consequence of there existing a \gwh spin-down connected with a  universal minimum ellipticity in \acp{MSP} of $\epsilon \approx 10^{-9}$. 
Reaching the level of sensitivity required to obtain a limit of $\epsilon \approx 10^{-9}$ for J0711\textminus6830 is not trivial with
second generation detectors. This would require the
planned network of five advanced detectors to reach their
design sensitivity \citep{2018LRR....21....3A}, and collect data for times exceeding at least 1-1.5
years of observation. 
On the other hand, that ellipticity level will be
accessible to 3rd generation detectors such as the Einstein Telescope \citep{Punturo_2010} and Cosmic Explorer \citep{2019BAAS...51g..35R}.

The \gwh data used here were drawn from the O1, O2, and O3 runs of the Advanced LIGO and Advanced Virgo detectors.  More data have been taken since, which will allow us to probe deeper still into the \gwh emission of spinning neutron stars.  Also, the analysis reported here involved five particularly interesting targets.  The full LIGO and Virgo data sets can be brought to bear on many more known pulsars.  Such an analysis is underway, and will be reported at a later date.

\section*{Acknowledgements}

The authors gratefully acknowledge the support of the United States
National Science Foundation (NSF) for the construction and operation of the
LIGO Laboratory and Advanced LIGO as well as the Science and Technology Facilities Council (STFC) of the
United Kingdom, the Max-Planck-Society (MPS), and the State of
Niedersachsen/Germany for support of the construction of Advanced LIGO
and construction and operation of the GEO600 detector.
Additional support for Advanced LIGO was provided by the Australian Research Council.
The authors gratefully acknowledge the Italian Istituto Nazionale di Fisica Nucleare (INFN),
the French Centre National de la Recherche Scientifique (CNRS) and
the Netherlands Organization for Scientific Research,
for the construction and operation of the Virgo detector
and the creation and support  of the EGO consortium.
The authors also gratefully acknowledge research support from these agencies as well as by
the Council of Scientific and Industrial Research of India,
the Department of Science and Technology, India,
the Science \& Engineering Research Board (SERB), India,
the Ministry of Human Resource Development, India,
the Spanish Agencia Estatal de Investigaci\'on,
the Vicepresid\`encia i Conselleria d'Innovaci\'o, Recerca i Turisme and the Conselleria d'Educaci\'o i Universitat del Govern de les Illes Balears,
the Conselleria d'Innovaci\'o, Universitats, Ci\`encia i Societat Digital de la Generalitat Valenciana and
the CERCA Programme Generalitat de Catalunya, Spain,
the National Science Centre of Poland,
the Swiss National Science Foundation (SNSF),
the Russian Foundation for Basic Research,
the Russian Science Foundation,
the European Commission,
the European Regional Development Funds (ERDF),
the Royal Society,
the Scottish Funding Council,
the Scottish Universities Physics Alliance,
the Hungarian Scientific Research Fund (OTKA),
the French Lyon Institute of Origins (LIO),
the Belgian Fonds de la Recherche Scientifique (FRS-FNRS),
Actions de Recherche Concertées (ARC) and
Fonds Wetenschappelijk Onderzoek – Vlaanderen (FWO), Belgium,
the Paris \^{I}le-de-France Region,
the National Research, Development and Innovation Office Hungary (NKFIH),
the National Research Foundation of Korea,
Industry Canada and the Province of Ontario through the Ministry of Economic Development and Innovation,
the Natural Science and Engineering Research Council Canada,
the Canadian Institute for Advanced Research,
the Brazilian Ministry of Science, Technology, Innovations, and Communications,
the International Center for Theoretical Physics South American Institute for Fundamental Research (ICTP-SAIFR),
the Research Grants Council of Hong Kong,
the National Natural Science Foundation of China (NSFC),
the Leverhulme Trust,
the Research Corporation,
the Ministry of Science and Technology (MOST), Taiwan
and
the Kavli Foundation.
The authors gratefully acknowledge the support of the NSF, STFC, INFN and CNRS for provision of computational resources.

We would like to thank all of the essential workers who put their health at risk during the COVID-19 pandemic, without whom we would not have been able to complete this work.


\software{The parameter estimation was performed with
  the \textsc{LALInference} \citep{PhysRevD.91.042003} library within
  \textsc{LALSuite} \citep{LALSuite}. All plots have been
  prepared using Matplotlib \citep{2007CSE.....9...90H}.}

\bibliography{references}

\end{document}